\documentclass[preprint,nofootinbib,11pt,floatfix,superscriptaddress,aps,prd,showpacs]{revtex4-2}
\usepackage{amsmath,amssymb,amsbsy,amsfonts,amsopn}
\usepackage{graphicx}
\usepackage{color,xcolor}
\usepackage{slashed,esint}

\usepackage[dvips]{epsfig}
\usepackage[dvips]{graphicx}
\usepackage{units}
\usepackage{textcomp}
\usepackage[normalem]{ulem}
\usepackage{hyperref}
\usepackage{placeins}
\newcommand{\beq}{\begin{equation}}
\newcommand{\eeq}{\end{equation}}
\newcommand{\bea}{\begin{eqnarray}}
\newcommand{\eea}{\end{eqnarray}}

\begin{document}

\title{\textbf{New Improved Schwarzschild Black Hole and Its Thermodynamics and Topological Classification}}

\author{G. Alencar}
\email{geova@fisica.ufc.br}
\affiliation{Departamento de F\'isica, Universidade Federal do Cear\'a, Caixa Postal 6030, Campus do Pici, 60455-760 Fortaleza, Cear\'a, Brazil.}

\author{T. M. Crispim} 
\email{tiago.crispim@fisica.ufc.br}
\affiliation{Departamento de F\'isica, Universidade Federal do Cear\'a, Caixa Postal 6030, Campus do Pici, 60455-760 Fortaleza, Cear\'a, Brazil.}

\author{C. R. Muniz}
\email{celio.muniz@uece.br}
\affiliation{Universidade Estadual do Cear\'a, Faculdade de Educa\c c\~ao, Ci\^encias e Letras de Iguatu, 63500-000, Iguatu, CE, Brazil.}

\author{M. Nilton}
\email{matheus.nilton@ufersa.edu.br}
\affiliation{Universidade Federal Rural do Semi-Árido, 59515-000, Campus Angicos, Angicos, Rio Grande do Norte, Brazil.}

\begin{abstract}

We construct a renormalization-group improved Schwarzschild-like black hole geometry using the exact new scheme running for the Newton coupling. The scale identification is implemented via a standard interpolating proper-distance function governed by an interpolation parameter $\gamma$ that smoothly connects the ultraviolet and infrared regimes. We present the resulting coordinate-dependent coupling and the improved metric function, analyzing its asymptotic expansions. The large-distance limit is shown to recover the classical Schwarzschild solution, while for $\gamma>0$ the short-distance behavior exhibits a regular de Sitter-like core, resulting in the regularization of the central singularity. We also analyze the thermodynamic properties of the solution, showing that quantum corrections significantly modify the small-radius behavior, leading to a remnant configuration and a nontrivial phase structure. Finally, we investigate the thermodynamic topology of the improved Schwarzschild black hole and show that the quantum-gravity corrections regularize the Hawking-temperature profile, giving rise to an additional thermodynamic defect with opposite winding number and changing the global thermodynamic-topological charge from the classical Schwarzschild value $W=-1$ to $W=0$.

\end{abstract}

\maketitle

\newpage

\section{Introduction}

Initially regarded as pure theoretical artifacts of General Relativity, black holes have emerged as pivotal laboratories for testing gravity in the strong-field regime. Recent years have witnessed an observational paradigm shift driven by the successful detection of gravitational waves from compact binary coalescences and the groundbreaking direct imaging of the M87* and Sgr A* black hole shadows \cite{LIGOScientific:2016aoc,EventHorizonTelescope:2019dse,EventHorizonTelescope:2022wkp,EventHorizonTelescope:2019ggy,EventHorizonTelescope:2019pgp}, paving the way for novel tests of spacetimes beyond the Kerr paradigm \cite{Duran-Cabaces:2025sly,Guerrero:2021ues,Olmo:2023lil,Olmo:2021piq,Olmo:2025ctf,GrallaHolzWald2019,Gulia:2025hce,Sekhmani:2025bsi,Ali:2025ney}.

Despite these observational triumphs, a fundamental problem within the classical black hole framework remains unresolved: the existence of singularities. In these spacetime regions, curvature invariants diverge, preventing geodesics from being extended and signaling a breakdown of physical predictability. Indeed, the celebrated singularity theorems proposed by Roger Penrose \cite{Penrose:1964wq} demonstrate that, under reasonable physical assumptions, the formation of a singularity is an inevitable outcome of gravitational collapse in classical General Relativity.

The emergence of these pathological features strongly indicates that standard General Relativity is incomplete at extreme energy scales. Over the years, various classical and phenomenological approaches have been extensively explored to regularize the spacetime geometry and circumvent central singularities, such as the construction of regular black hole models and modified gravity frameworks \cite{Bardeen:1968,Ayon-Beato:1998hmi,Bronnikov:2000vy,Bronnikov:2005gm,Hayward:2005gi,Sotiriou:2008rp,Olmo:2015bya,Frolov:2016pav,Alencar:2024yvh,Alencar:2025jvl,Lima:2023jtl,Silva:2025fqj,Lobo:2020ffi,Rodrigues:2023vtm,deSousaSilva:2018kkt,Junior:2020zdt,Nojiri:2010wj,Nojiri:2017ncd,Nojiri:2017kex,Battista:2023iyu,Wang:2025fmz}. However, a fundamental and complete resolution of these divergences is generally expected to require taking quantum gravity effects into account at the Planck scale.

Renormalization-group (RG) improvement of classical spacetimes represents a powerful phenomenological approach to incorporate quantum gravity effects into classical gravitational solutions. This method leverages the renormalization group flow of gravitational couplings to modify classical metrics, capturing key quantum corrections while maintaining computational tractability. Within the framework of Asymptotic Safety, gravity is proposed to be non-perturbatively renormalizable through the existence of an ultraviolet (UV) fixed point that governs the high-energy behavior of the theory \cite{Reuter:1996cp,Niedermaier:2006wt,Reuter:2019byg}. This formalism has been broadly applied to a wide range of gravitational scenarios (See Refs.\cite{Saueressig:2015xua,Pawlowski:2018swz,Platania:2019kyx,Bosma:2019aiu,Ishibashi:2021kmf,Chen:2022xjk,Koch:2016uso,Rincon:2017ypd,Rincon:2017goj,Rincon:2018sgd,Rincon:2018dsq,Rincon:2018lyd,Rincon:2019zxk,Fathi:2019jid,Rincon:2022hpy,Falls:2018ylp,Draper:2020bop,Platania:2020knd,Fehre:2021eob,Alencar:2021mus,Nilton:2022cho,Nilton:2022hrp,Nilton:2022jji,Donmez:2026dfv,Falls:2010he,Falls:2012nd,Litim:2013gga,Kluth:2020bdv,Kluth:2022vnq,Falkenberg:1996bq,Nojiri:1996vs}). 

In this program, the Newton coupling $G$ exhibits a characteristic running with the RG cutoff scale $k$, following a flow equation derived from the functional renormalization group \cite{Reuter:1996cp,Litim:2003vp}. The fundamental step in implementing RG improvement involves identifying this cutoff scale with a physical inverse length scale proportional to $\sim 1/d(r)$, where $d(r)$ represents an appropriate distance measure in the spacetime. This identification procedure transforms the scale-dependent coupling $G(k)$ into a coordinate-space effective coupling $G(r)$, ultimately yielding RG-improved metrics that interpolate between quantum and classical regimes \cite{Bonanno:2000ep}.

Over the past two decades, RG-improved black hole solutions have been extensively investigated within the asymptotic-safety program. Beyond the original construction of the improved Schwarzschild solution \cite{Bonanno:2000ep}, several generalizations have been explored, including charged and rotating geometries, (anti-)de Sitter backgrounds, and extensions incorporating matter fields and higher-derivative operators \cite{Bonanno:2006eu,Falls:2017lst,Koch:2014cqa,Platania:2019kyx,Rincon:2018lyd}. These studies consistently indicate that the scale dependence of the gravitational couplings can soften or remove classical singularities, modify the near-horizon structure, and generate nontrivial phenomenological implications at Planckian scales. 

In particular, the thermodynamic properties, but not their topological signatures, of RG-improved Schwarzschild black holes based on the standard Bonanno--Reuter (``Scheme-C'') running were investigated in Ref.~\cite{Mandal:2022quv}, while alternative asymptotic-safety-inspired regular black hole geometries have been obtained from dust-collapse constructions, leading to distinct logarithmic lapse functions and thermodynamic behavior \cite{Zhang:2024zqd}. More recently, the novel exact ``Scheme-B'' running was considered in Ref.~\cite{Muniz:2026drk}, where its infrared sector was analyzed in the context of RG-improved black holes. However, due to the absence of a complete UV/IR interpolation, the resulting geometry remained singular. The present work extends this program by implementing the exact Scheme-B running throughout the full UV/IR domain, yielding a regular RG-improved Schwarzschild solution and enabling a comprehensive geometric, thermodynamic, and topological analysis.

The choice of the distance function $d(r)$ constitutes a crucial aspect of this construction, as it determines how quantum corrections manifest across different spacetime regions. In the present work, we employ the interpolating proper-distance function
\begin{equation}
    d(r) = \left( \frac{r^3}{r + \gamma G_0 M} \right)^{1/2},
\end{equation}
which has been extensively used in the literature \cite{Bonanno:2000ep,Bonanno:2006eu} due to its desirable asymptotic properties. This specific form ensures that the cutoff identification $\Lambda(r) = \xi/d(r)$ produces a smooth transition between the infrared (IR) and ultraviolet (UV) regimes of the geometry, effectively capturing the scale-dependent nature of quantum gravitational effects.

The physical interpretation of this construction becomes transparent in the asymptotic limits. In the ultraviolet regime for $\gamma>0$, $r \ll \gamma G_0 M$, corresponding to scales deep within the quantum-dominated region, the distance function simplifies to
\begin{equation}
 d(r) \simeq \frac{r^{3/2}}{\sqrt{\gamma G_0 M}},
 \qquad \Rightarrow \qquad
 \Lambda(r) \simeq r^{-3/2}.
\end{equation}

The dimensionless interpolation parameter $\gamma$ controls the scale at which quantum gravitational effects become dominant. The case $\gamma<0$ must be excluded because the denominator of $d(r)$ vanishes at the positive radius $r=|\gamma|G_0M$, causing the distance function to diverge and rendering the cutoff identification $\Lambda(r)$ pathological. On the other hand, $\gamma=0$ is mathematically well-defined for $r>0$, yielding
\begin{equation}
d(r)=r,
\end{equation}
so that the cutoff distance remains real and positive. However, in this case the interpolating ultraviolet regime is absent, and the scaling responsible for the regular de Sitter-like core is not obtained. Therefore, while $\gamma=0$ is admissible from the standpoint of the cutoff function itself, the condition $\gamma>0$ is required in the present construction to generate the regular de Sitter-like core and consequently to regularize the central curvature singularity.

For $\gamma>0$, this enhanced scaling behavior amplifies the running of the gravitational coupling at small distances, activating the asymptotic safety corrections that fundamentally alter the spacetime structure. Consequently, within this effective framework, this UV regime regularizes the curvature singularity, replacing it with a de Sitter-like core
\cite{Bonanno:2000ep,Falls:2017lst}.

Conversely, in the infrared limit $r \gg \gamma G_0 M$, representing the classical asymptotic region, the distance function approaches the areal radius, $d(r) \to r$, leading to the cutoff scaling $\Lambda(r) \sim 1/r$. In this regime, the running coupling asymptotically approaches its classical value, $G(r) \to G_0$, and the improved metric reduces to the standard Schwarzschild solution, thus ensuring the recovery of general relativity at large distances. The interpolating nature of $d(r)$ therefore guarantees a physically consistent transition between the quantum gravity dominated core, governed by the UV fixed point, and the classical exterior region where Einstein's theory remains valid.

The core of this research is centered on the exact ``Scheme B'' formulation of the dimensionful running Newton coupling, as recently introduced in \cite{Bonanno:2025dry}. In the limit where the cosmological constant vanishes ($\Lambda_0 = 0$), this framework provides an analytically tractable scenario to explore how Asymptotic Safety corrections reshape the fundamental laws of black hole thermodynamics. The scale dependence $G(r)$ redefines the relationship between the black hole mass and the horizon radius, directly impacting the Hawking temperature $T_H$ and the heat capacity $C$. Unlike the classical Schwarzschild behavior, where the temperature diverges as the mass vanishes and the heat capacity remains strictly negative, RG-improved corrections introduce a non-trivial phase structure. This allows for the existence of stable remnants ($C > 0$) and fundamentally alters the evaporation dynamics at the Planck scale.

This restructuring of the thermodynamic landscape motivates an investigation that transcends local analysis, necessitating a global topological classification of the phase space. Utilizing the generalized free energy formalism and its associated vector field $\phi$ \cite{Wei:2022dzw,Wu:2024asq,Yang:2025uul,Wu:2025xxo}, we treat thermodynamic equilibrium points as topological defects. The modification of the metric by the RG flow can alter the number and nature of these critical points, which is manifested as shifts in the winding numbers and the global topological number $W$. Such an approach allows us to discern whether quantum gravity corrections merely deform the classical solution or induce a qualitative, topological transition in the system's thermodynamic structure.

This work is organized as follows: In Section \ref{sII}, we detail the UV/IR interpolation within Scheme B and the identification of the cutoff scale. In Section \ref{sIII}, we derive the new improved Schwarzschild metric and examine its geometric properties and asymptotic limits. Section \ref{sIV} is dedicated to the investigation of the thermodynamic properties of the solution, while Section \ref{sV} presents its topological analysis in the phase space. Finally, we present our Conclusions in Section \ref{sVI}. Throughout this work, we adopt the metric signature $(-, +, +, +)$ and work in natural units where $c = \hbar = 1$, while the Newton constant $G$ remains scale-dependent.


\section{Interpolation UV/IR in Scheme-B}\label{sII}

In the RG-improvement approach, quantum gravity effects are incorporated through the running of the gravitational couplings and a suitable UV/IR scale identification. As we have already indicated, here we employ the standard proper-distance interpolation, which promotes the running Newton coupling to a coordinate-dependent quantity and smoothly connects the quantum ultraviolet regime to the classical infrared limit. The resulting metric should be interpreted as an effective phenomenological geometry rather than as a solution derived from a fully specified effective gravitational action. A derivation from the effective average action would generally generate modified field equations containing additional derivative contributions associated with the spacetime variation of $G(r)$. The present work focuses instead on the consequences of the exact Scheme-B running within the standard RG-improvement framework.

Thus, we take the scheme-B expression for the Newton coupling in the limit \(\Lambda_0=0\). Denoting the IR Newton constant by \(G_0\) and the dimensionless fixed-point value by \(g_*\), the running coupling (with \(\Lambda_0=0\)) reads \cite{Bonanno:2025dry}
\begin{equation}
G^{(B)}(\Lambda)\Big|_{\Lambda_0=0}
=
\frac{G_0}{
\displaystyle
\frac{\Lambda^2}{2}\!\left(\frac{G_0}{g_*}\right)
+
\sqrt{\,1 \;+\; \frac{1}{4}\,\Lambda^4\!\left(\frac{G_0}{g_*}\right)^{\!2}\,}
}\,.
\label{eq:G_B_L0}
\end{equation}

Following the interpolating prescription of Bonanno \& Reuter \cite{Bonanno:2000ep}, we identify the RG cutoff with an inverse proper-like distance:
\begin{equation}
\Lambda(r)\left(\frac{G_0}{g_*}\right)^{1/2}=\frac{\xi}{d(r)},\qquad 
d(r)=\left(\frac{r^3}{r+\gamma G_0 M}\right)^{1/2},
\label{eq:Lambda_d}
\end{equation}
where we introduced a cutoff distance scale  $\xi$, with $\gamma$ being the interpolation parameter. We emphasize that the cutoff identification employed here is a phenomenological prescription rather than a first-principles derivation. Because there is no unique method for mapping the renormalization group momentum scale to coordinate space—as alternative approaches might employ curvature invariants such as the Kretschmann scalar—the parameter $\xi$ effectively absorbs the theoretical ambiguity inherent to this mapping. Consequently, $\xi$ is not a fundamental coupling, but a scale-matching parameter linking $d(r)$ to $\Lambda(r)$. Although its physical magnitude is expected to be of the order of the Planck scale, we treat it here as a free parameter to systematically explore the model's geometric and thermodynamic sensitivity.

Substituting (\ref{eq:Lambda_d}) into \eqref{eq:G_B_L0} yields the coordinate-space coupling
\begin{equation}
G(r)=
\frac{G_0}{
\displaystyle
\frac{\xi^2}{2 d(r)^2}
+
\sqrt{\,1 \;+\; \frac{\xi^4}{4 d(r)^4}\,}
}\,,
\label{eq:G_of_r}
\end{equation}
with \(d(r)\) given by \eqref{eq:Lambda_d}.

\section{New Improved Schwarzschild Metric}\label{sIII}

Replacing \(G_0\) by \(G(r)\) in the Schwarzschild lapse we obtain the RG-improved metric
\begin{equation}
ds^2 = -f(r)\,dt^2 + f(r)^{-1}\,dr^2 + r^2 d\Omega^2,
\end{equation}

where the metric potential is
\begin{equation}
f(r)=1-\frac{2M\,G(r)}{r}=1-\frac{4 M r^2}{\xi ^2(\gamma  M + r)+\sqrt{\xi ^4 (\gamma  M+r)^2+4 r^6}}
\label{eq:metric}
\end{equation}
with \(G(r)\) as in \eqref{eq:G_of_r}. This is the explicit metric constructed with the exact scheme-B coupling (in the \(\Lambda_0=0\) limit) and the interpolating cutoff identification \(\Lambda(r)\sim\xi/d(r)\).

\begin{figure}[hb!]
    \centering 
    \includegraphics[width=0.495\textwidth]{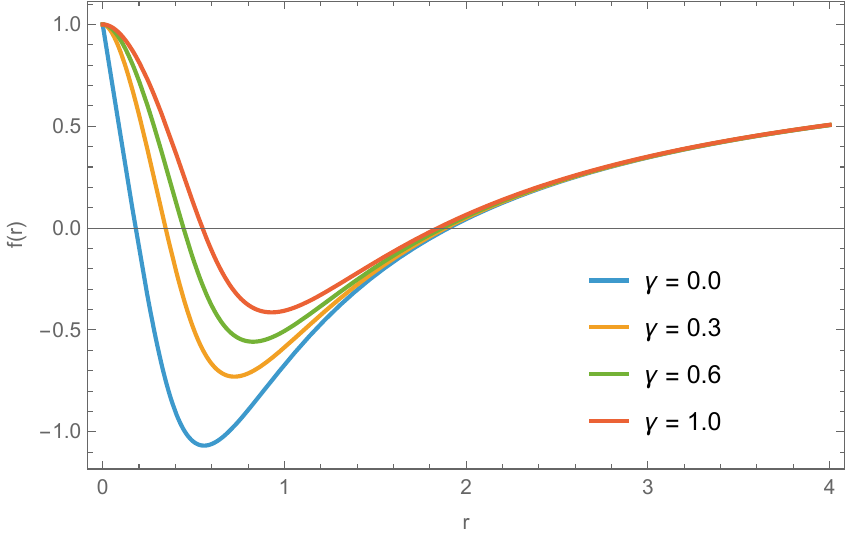}
    \includegraphics[width=0.495\textwidth]{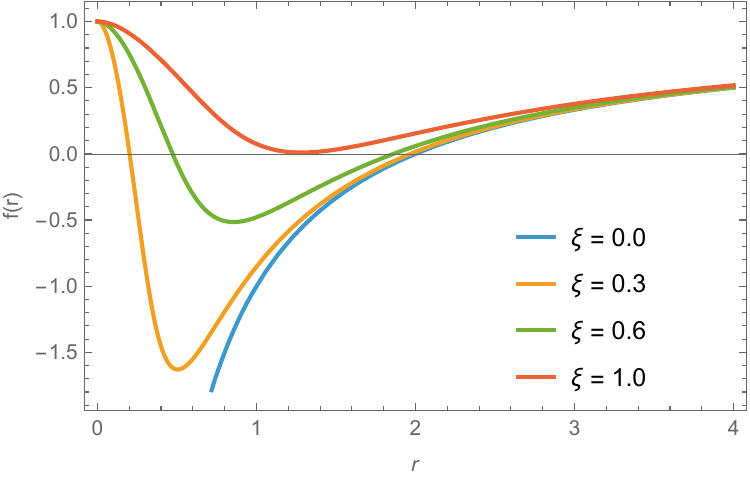}
    \caption{Metric potential as a function of the radial coordinate, for $M=1.0$. Left panel: Varying the interpolation parameter $\gamma$, with $\xi=0.6$. Right panel: Varying the cutoff scale $\xi$, with $\gamma=0.7$. Note that the $\xi=0.0$ curve represents the Schwarzschild black hole.}
    \label{fig:f(r)}
\end{figure}

\begin{figure}[h!]
    \centering 
    \includegraphics[width=0.4\textwidth]{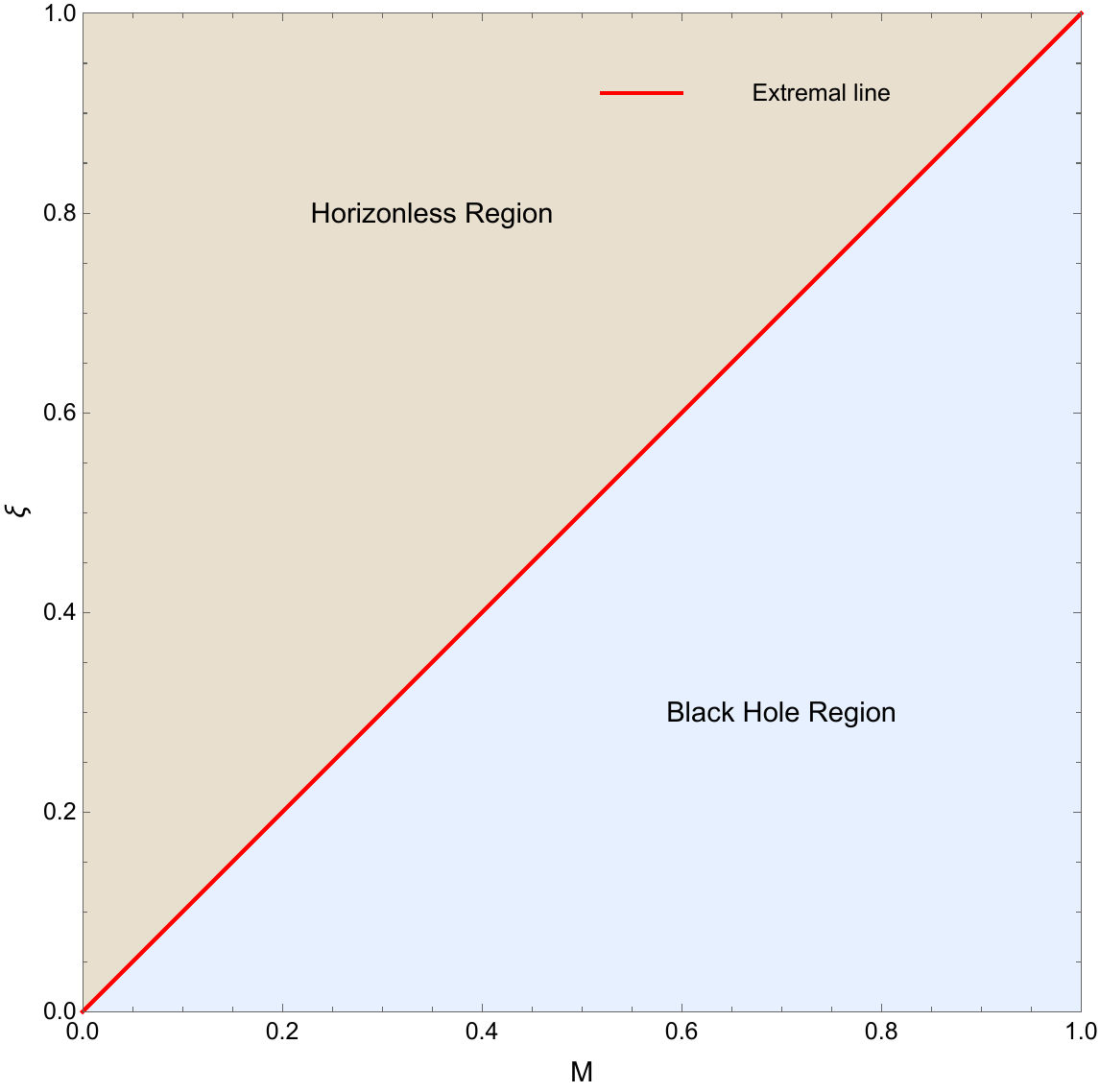}
    \caption{Parameter space $(M,\xi)$ of the new improved Schwarzschild black hole. The shaded region indicates the values of the parameters yielding black hole solutions with event horizons, bounded by the critical line where the horizon degenerates. Beyond this boundary, the spacetime describes no black hole. The interpolation parameter is set to $\gamma=0.6$.}
    \label{fig:param_space}
\end{figure}

The parameter space $(M, \xi)$ of the RG-improved Schwarzschild geometry is illustrated in Fig.~\ref{fig:param_space}. The shaded blue region denotes the domain where the lapse function $f(r)$ admits at least one real, positive root, signaling the presence of an event horizon. The boundary of this region, represented by the solid red line, corresponds to the extremal limit where the inner and outer horizons coincide. Due to the scaling properties of the metric, this extremal boundary is a straight line passing through the origin.

To understand the physical significance of this threshold, it is instructive to revisit the relation $\xi^2 \sim G_0/g_*$ derived from our scale identification, Eq. (\ref{eq:Lambda_d}). This relation establishes $\xi$ as a measure of the tension between macroscopic classical gravity, governed by $G_0$, and the dimensionless quantum ultraviolet fixed point, $g_*$. As $\xi$ increases, the regularizing quantum gravity effects dictated by $g_*$ become dominant at larger radial distances. Consequently, within this cutoff prescription, the extremal line marks the threshold where the effective geometry ceases to admit horizons. For any given mass $M > 0$, exceeding this critical cutoff value entirely prevents event horizon formation, resulting in a regular horizonless configuration instead.

To analyze the IR limit directly in the metric, we expand \(f(r)\) for large \(r\). For \(r\gg \gamma M\) we have \(d(r)\simeq r\). Hence the lapse expands as
\begin{equation}
f(r)\simeq 1 - \frac{2M }{r} + \frac{M \xi^2}{r^3} + O(r^{-4}).
\label{eq:f_IR}
\end{equation}

The metric therefore recovers the classical Schwarzschild form \(1-2M/r\) at leading order.

The UV limit is
\begin{equation}\label{metric-origin}
  f(r)\approx  1 - \frac{2 r^2}{\gamma \xi^2}+\mathcal{O}(r^3),
\end{equation}
which turns the solution de Sitter-like and therefore regular at the origin. 
\begin{figure}[h!]
    \centering 
    \includegraphics[width=0.495\textwidth]{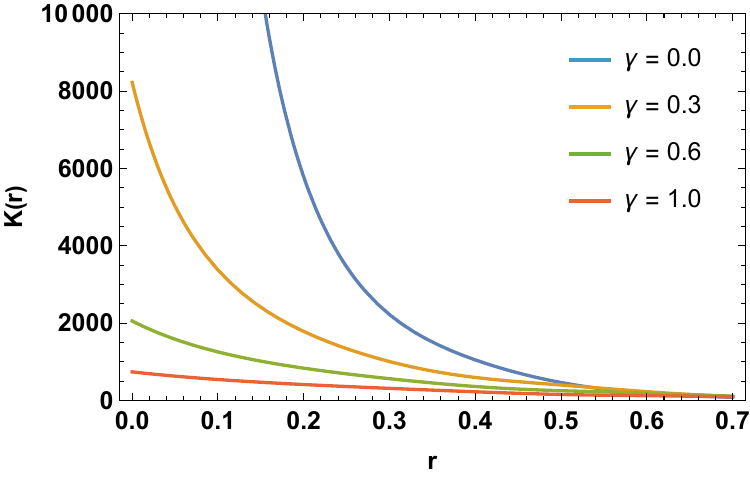}
    \includegraphics[width=0.495\textwidth]{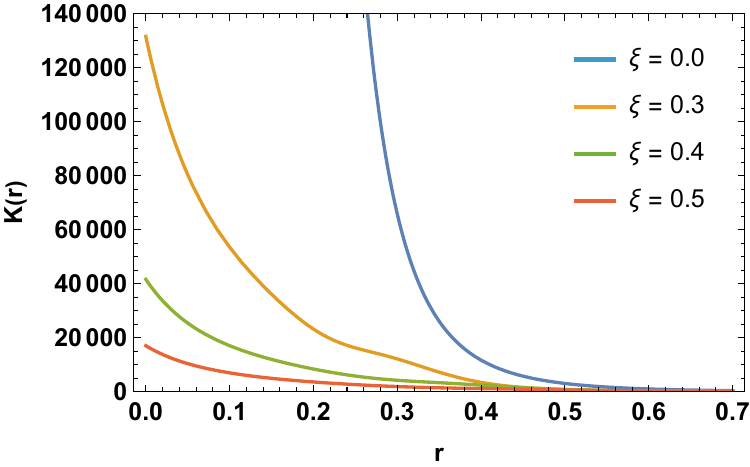}
    \caption{Left panel: Kretschmann scalar as a function of $r$, varying the interpolation parameter $\gamma$, with $\xi=0.6$. Right panel: The same quantity, now varying the cutoff scale $\xi$, with $\gamma=0.3$. We set  $M=1.0$.}
    \label{Kretschmann}
\end{figure}

We now turn to the analysis of the curvature properties of the spacetime under inspection, focusing on the Kretschmann scalar \(K(r)\). This invariant, which measures the magnitude of the Riemann tensor, is given by
\begin{equation}
K(r) = R_{\mu\nu\sigma\lambda}R^{\mu\nu\sigma\lambda}
      = \big(f''(r)\big)^{2}
      + \frac{4}{r^{2}}\big(f'(r)\big)^{2}
      + \frac{4}{r^{4}}\big(1 - f(r)\big)^{2}.
      \label{Kretsch}
\end{equation}

It provides a convenient diagnostic of the spacetime curvature and allows us to test whether the quantum-corrected geometry remains regular at the origin.
Using the asymptotic expansion of Eq.~\eqref{Kretsch}, the leading infrared behavior of the Kretschmann scalar \(K\) is
\begin{align}
K(r) &= \frac{48\,M ^2}{r^6} + O(r^{-8}),
\label{eq:K_IR}
\end{align}
with the leading term of \(K\) being identical to the Schwarzschild behavior, consistent with \(G(r)\to G_0\) as \(r\to\infty\).  On the other hand, the ultraviolet expansion (small $r$) for the metric given in Eq. (\ref{metric-origin}) leads to
\begin{equation}
    K=\frac{96}{\gamma ^2 \xi ^4}+\mathcal{O}(r),
\end{equation}
for $\xi\neq0$. This confirms that the spacetime is regular at the origin for nonzero values of the interpolation parameter $\gamma$, for which the RG-improved geometry develops a de Sitter-like core with finite curvature invariants. In contrast, for $\gamma=0$ the regularization mechanism is absent and the classical curvature singularity remains. This behavior is illustrated in Fig.~\ref{Kretschmann}, where the Kretschmann scalar is displayed as a function of the radial coordinate. In the left panel, the curve corresponding to $\gamma=0$ diverges as $r\to0$, while the cases with $\gamma>0$ remain finite. In the right panel, the curve with $\xi=0$ reproduces the Schwarzschild solution, whereas nonzero values of $\xi$ soften the ultraviolet behavior of the curvature invariant.

\section{Thermodynamics}\label{sIV}

In this section, we investigate the thermodynamic properties of the new improved Schwarzschild solution characterized by the lapse function in Eq.\eqref{eq:metric}. In particular, we examine how the cutoff scaling parameter $\xi$ and the interpolation parameter $\gamma$ modify the thermodynamic behavior and influence the topological classification of the solution.

We first compute the Hawking temperature associated with the improved Schwarzschild solution. For a static and spherically symmetric spacetime, the Hawking temperature is given by
\begin{equation}
\label{eq:temperature-def}
T_{\mathrm{H}}(r_+)=\frac{1}{4\pi}\left.\frac{d f(r)}{dr}\right|_{r=r+},
\end{equation}
where $r_+$ denotes the radial position of the event horizon, determined by the condition $f(r_+)=0$. Imposing this condition leads to the relation
\begin{equation}
\label{eq:horizon-condition}
4Mr_+^2 - \xi^2(\gamma M + r_+) = \sqrt{\xi^4 (\gamma M + r_+)^2 + 4 r_+^6},
\end{equation}
which can be solved for the mass parameter $M$ as a function of the horizon radius. By squaring both sides of Eq. \eqref{eq:horizon-condition} and simplifying, we obtain a quadratic equation for the mass:
\begin{equation}
    (4r_+^2 - 2\gamma\xi^2)M^2 - (2\xi^2 r_+)M - r_+^4 = 0.
\end{equation}
Solving this equation yields two distinct mathematical branches for $M$. The branch corresponding to the negative sign in front of the discriminant leads to a strictly negative mass parameter. Furthermore, in the classical limit $\xi \to 0$, this negative branch fails to recover the standard Schwarzschild relation $M = r_+/2$. Therefore, it must be discarded as physically inadmissible. The physically relevant solution is uniquely given by the positive branch, which, after algebraic simplification, takes the compact form\begin{equation}\label{eq:mass}M(r_+)=\frac{r_+^3}{-\xi^2+\sqrt{4r_+^4-2\gamma\xi^2 r_+^2+\xi^4}}.\end{equation}The resulting expression is well-defined only within a restricted region of the parameter space. Requiring the positivity of the mass parameter implies the condition $\sqrt{4r_+^4-2\gamma\xi^2 r_+^2+\xi^4} > \xi^2$, which leads directly to the constraint on the horizon radius:
\begin{equation}
\label{constraint}
4 r_+^2 - \xi^2 \gamma > 0 \rightarrow r_+ > \xi\sqrt{\gamma/2}.
\end{equation}
 This condition constrains the allowed values of $r_+$ for which an event horizon exists, thereby delimiting the physically admissible sector of the solution.

From Eq.\eqref{eq:temperature-def}, the Hawking temperature associated with the improved Schwarzschild solution can be written as
\begin{equation}
\label{eq:temperature}
T_{\textrm{H}}(r_+)=\frac{1}{4\pi}\left[-\frac{2}{r_+}+\frac{1}{4r_+^2 M(r_+)}\left(\xi^2 + \frac{\xi^4(\gamma M(r_+)+r_+)+12r_+^5}{\sqrt{\xi^4(\gamma M(r_+)+r_+)^2+4r_+^6}}\right)\right],
\end{equation}
where $M(r_+)$ denotes the mass parameter expressed in terms of the horizon radius $r_+$, Eq.\eqref{eq:mass}. 

\begin{figure}[h!]
    \centering 
    \includegraphics[width=0.60\textwidth]{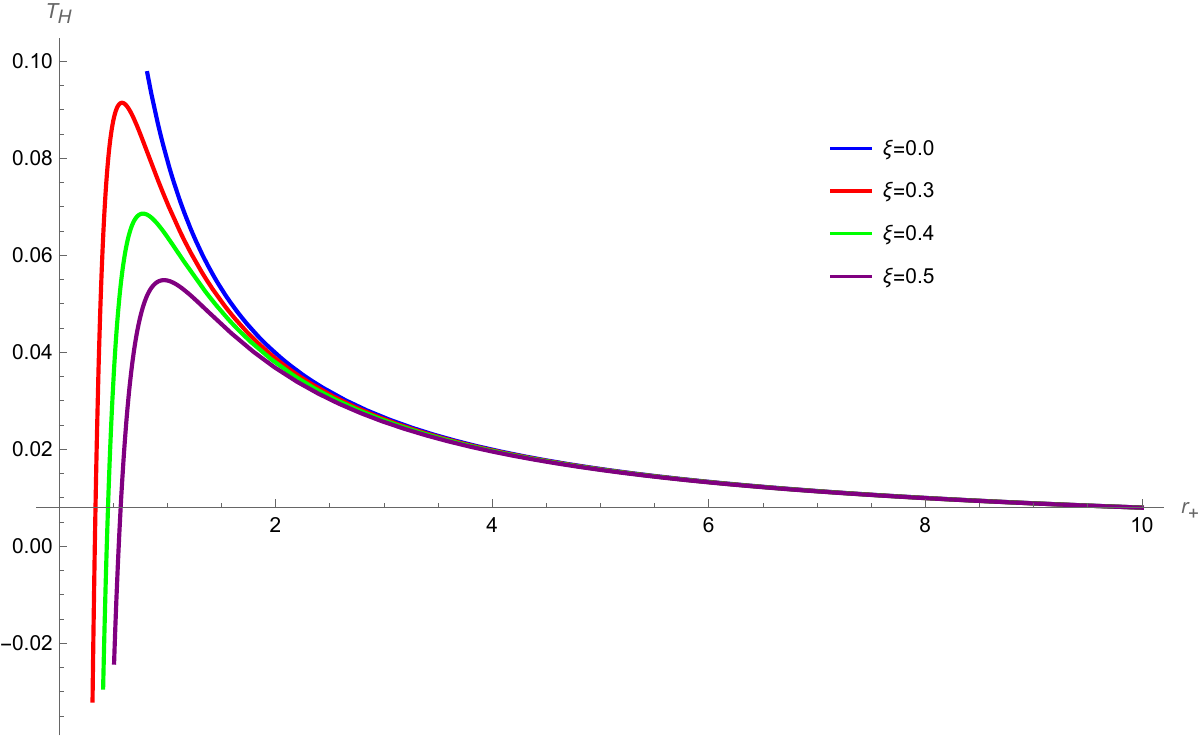 }
    \includegraphics[width=0.60\textwidth]{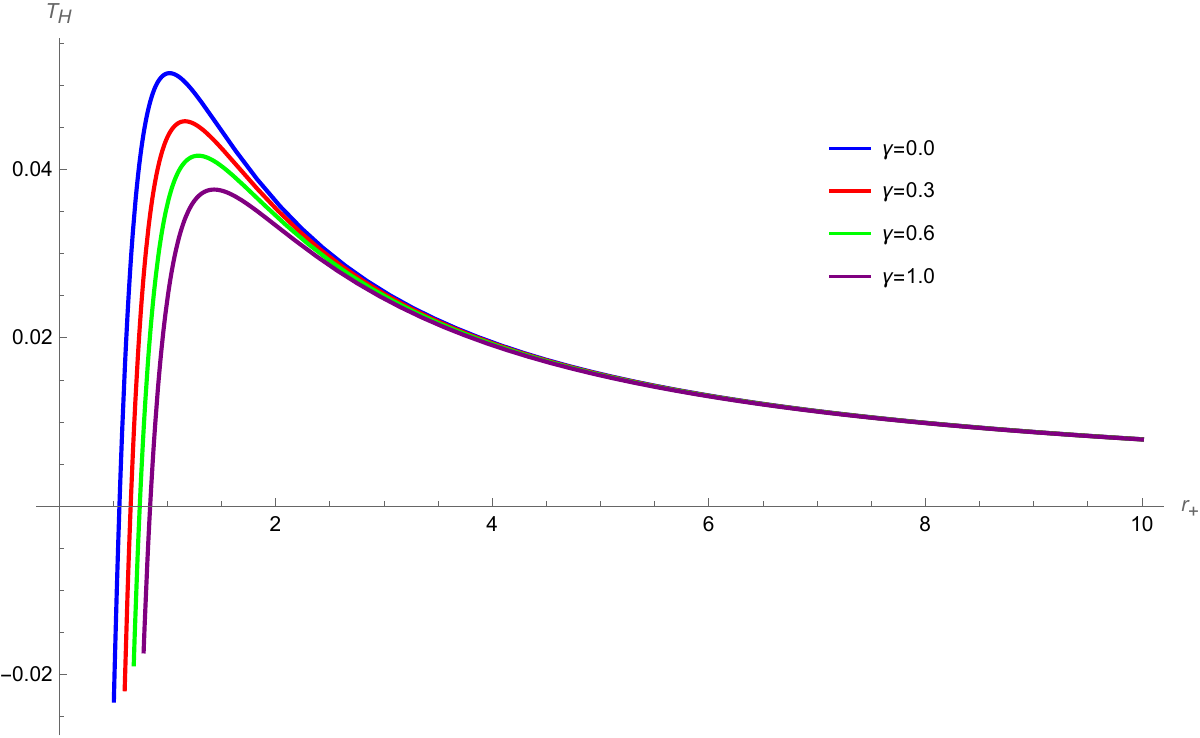}
    \caption{ Top panel: Hawking temperature as a function of the horizon radius $r_+$, varying the cutoff scale $\xi$, with $\gamma=0.3$. Bottom panel: The same quantity, now varying the interpolation parameter $\gamma$, with $\xi=0.6$.}
    \label{fig:temperature}
\end{figure}

It is worth emphasizing that, in the classical limit $\xi\rightarrow0$, the above expression for temperature smoothly reduces to the standard Hawking temperature of the Schwarzschild black hole. The behavior of the Hawking temperature for different values of the parameters $\xi$ and $\gamma$ can be seen in Fig.\ref{fig:temperature}. 

We observe that in the IR regime, corresponding to large values of the horizon radius, the Hawking temperature reproduces the standard Schwarzschild behavior, as expected. This confirms that the asymptotically safe gravity (ASG) corrections do not affect the classical large-scale limit of the solution. Significant deviations from the classical behavior arise only for small horizon radii, where quantum gravity-inspired corrections become relevant. In particular, while the classical Schwarzschild temperature diverges as $r_+\rightarrow0$, the presence of the ASG cutoff parameter $\xi$ regularizes this divergence, leading to a smooth behavior of the temperature in the UV regime.

A notable feature of the improved solution is the emergence of a maximum temperature at a finite value of $r_+$.
Below this point, the temperature decreases smoothly and eventually vanishes at a finite radius, signaling the breakdown of the semiclassical description and suggesting the possible formation of a cold remnant. Such behavior is absent in the classical Schwarzschild case. Furthermore, we find that increasing the values of the parameters lowers the maximum temperature attained by the black hole. This indicates that stronger quantum gravity effects enhance the regularizing behavior of the solution, reinforcing the suppression of ultraviolet divergences in the thermodynamic quantities.

Now, let us turn to the entropy of the black hole. In theories of modified gravity, the presence of new parameters often raises the question of whether the first law of thermodynamics requires extended work terms. In our framework, we treat the cutoff scaling parameter $\xi$ and the interpolation parameter $\gamma$ as fixed coupling constants of the underlying theory rather than dynamic thermodynamic variables. Under this strict assumption, the standard first law of black hole thermodynamics, $dM = T_H dS$, remains entirely valid without additional variations. Consequently, the entropy can be consistently computed from the first law through the relation
\begin{equation}\label{eq:entropy-def}
S(r_+) = \int^{r_+} \frac{dM}{T_H}
       = \int^{r_+} \frac{1}{T_H}\,\frac{dM}{dr_+}\,dr_+ ,
\end{equation}
where the mass function $M(r_+)$ and the Hawking temperature $T_H(r_+)$ are given by Eqs.~\eqref{eq:mass} and~\eqref{eq:temperature}, respectively. 
The integral in Eq.~\eqref{eq:entropy-def} is highly nontrivial and does not admit a closed analytic expression in terms of elementary functions. It should also be emphasized that Eq.~(\ref{eq:entropy-def}) is defined up to an arbitrary additive integration constant, which here will be made equal to zero. Obviously, different choices of this constant modify the absolute normalization of the entropy and of the generalized free energy. However, since the thermodynamic and topological analyses performed in this work depend only on derivatives of these quantities, the critical behavior and the resulting physical conclusions are unaffected.

As we have seen, the parameter $\xi$ encodes the strength of asymptotically safe quantum gravitational corrections. In the limit $\xi \to 0$, the classical Schwarzschild solution is smoothly recovered. Therefore, in order to gain analytic insight into the thermodynamic behavior of the solution, it is natural to consider the regime $\xi \ll r_+$, which corresponds to small quantum corrections around the classical geometry. In this regime, the entropy can be consistently approximated by performing a perturbative expansion of the integrand in powers of $\xi$.

Expanding the integrand of Eq.~\eqref{eq:entropy-def} around $\xi = 0$ and integrating term by term, we obtain the following approximate expression for the entropy:
\begin{equation}\label{eq:entropy}
S(r_+) = \pi r_+^2 + \pi(1+\gamma)\,\xi^2 \ln r_+ + \mathcal{O}(\xi^4).
\end{equation}
As expected, the classical Bekenstein--Hawking area law, $S = \pi r_+^2$, is recovered in the limit $\xi \to 0$. 
The leading quantum correction appears as a logarithmic term in the horizon radius. Logarithmic corrections to black hole entropy are a well-known and robust feature of quantum gravity, arising in a wide variety of approaches, including the generalized uncertainty principle (GUP), loop quantum gravity, and quantum field theory on curved spacetimes \cite{Medved:2005vw,Chen:2009sp, Banerjee:2010qc,Sen:2012dw,Keeler:2014bra,Nozari:2006vn,Faizal:2014tea,Anacleto:2020zfh}. 

\begin{figure}[!htpb]
    \centering
    \includegraphics[width=0.675\linewidth]{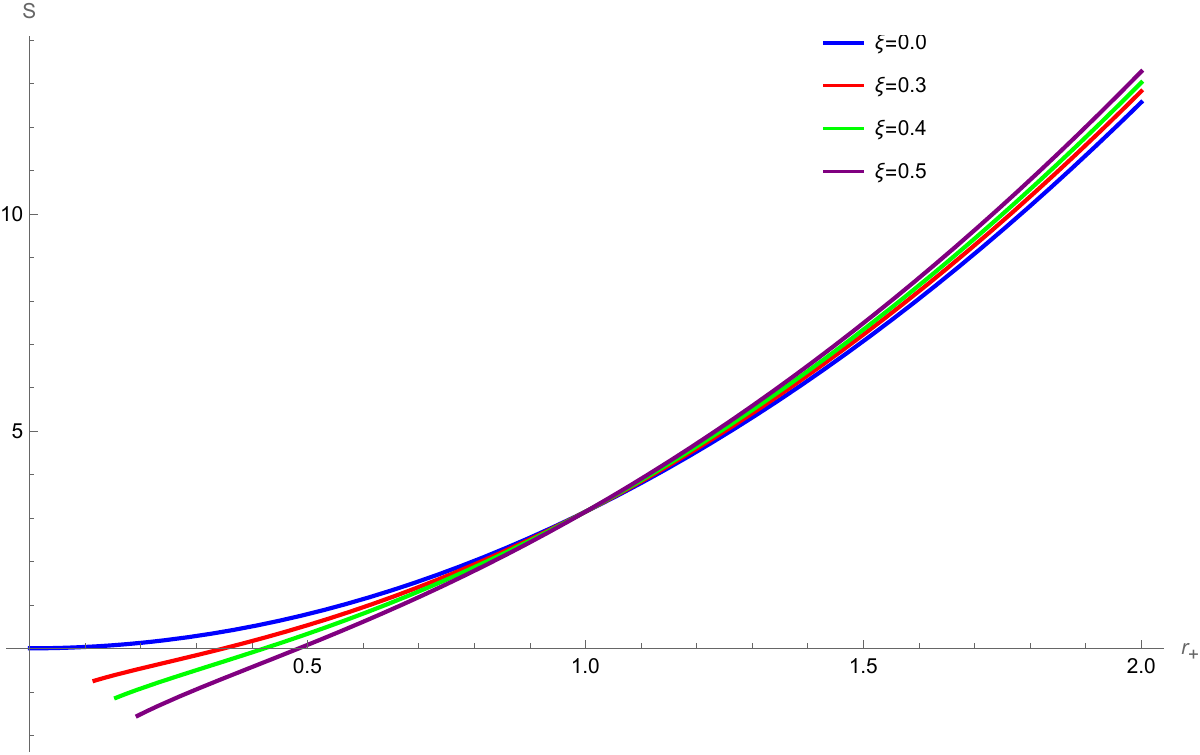}
    \caption{Approximate black hole entropy obtained from a perturbative expansion up to second order in $\xi$, for different values of the parameter $\xi$. In all curves we fix $\gamma = 0.3$.}
    \label{fig:entropy}
\end{figure}

It is worth emphasizing that the coefficient of the logarithmic correction depends explicitly on the parameter $\gamma$, reflecting the model-dependent nature of the quantum gravitational effects. While the logarithmic functional form itself is a familiar feature shared by various quantum gravity approaches, its specific derivation here relies heavily on the perturbative regime $\xi \ll r_+$ and the chosen distance identification scheme. Therefore, rather than being strictly universal, this modification is intrinsically tied to the macroscopic modeling of the RG improvement and the employed expansion. The behavior of the approximate expression for entropy, for different values of the parameter $\xi$, can be seen in Fig.\ref{fig:entropy}. A particularly interesting feature of Eq.~\eqref{eq:entropy} is the dependence of the logarithmic prefactor on the quantity $(1+\gamma)$. Mathematically, one might notice that the special choice $\gamma = -1$ would exactly cancel this leading-order quantum correction, restoring the purely classical area law at this order. However, as previously discussed, this value falls outside the physically admissible domain of the model. A negative interpolation parameter $\gamma$ introduces a divergence in the proper-distance function $d(r)$ at finite radii, breaking down the geometric regularization scheme long before the central core is reached. Consequently, within the physically consistent parameter space ($\gamma > 0$), the logarithmic entropy correction is strictly non-vanishing, reaffirming its robust nature within the asymptotically safe gravity framework.

\begin{figure}[!htpb]
    \centering
    \includegraphics[width=0.65\linewidth]{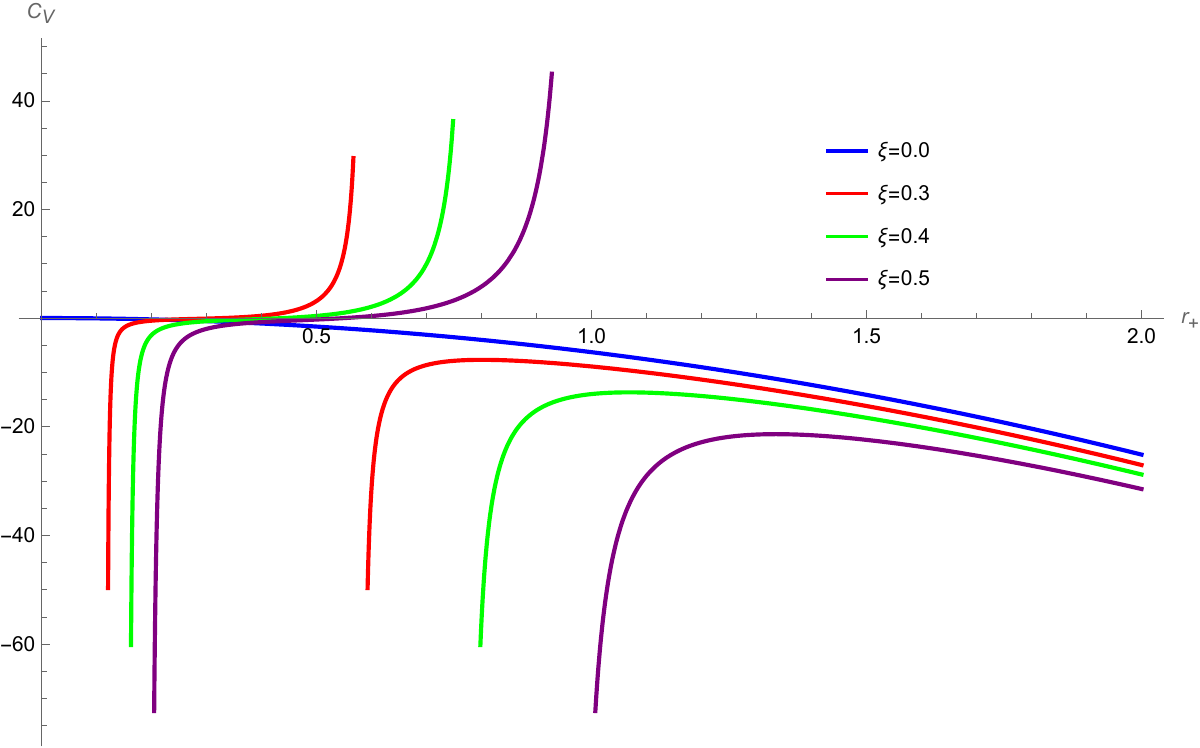}
    \caption{Heat capacity at constant volume for different values of $\xi$. In all curves we have considered $\gamma=0.3$.}
    \label{fig:heat-capacity}
\end{figure}

Finally, let us now consider the constant volume heat capacity, whose definition is given by
\begin{equation}\label{eq:heat-capacity-def}
C_V(r_+)=\frac{\partial M}{\partial T_H}=\frac{\partial M}{\partial r_+}\left(\frac{\partial T_H}{\partial r_+}\right)^{-1}.
\end{equation}
The full expression for the heat capacity can be obtained straightforwardly from the mass and Hawking temperature, given by Eqs.~\eqref{eq:mass} and \eqref{eq:temperature}. However, the resulting formula is too lengthy to be displayed explicitly. Therefore, we focus on a graphical analysis to highlight the effects of the quantum corrections associated with asymptotically safe gravity, which are encoded in the parameter $\xi$. The behavior of the heat capacity for various values of the parameter $\xi$ can be seen in Fig.~\ref{fig:heat-capacity}.

We can see directly that the classical Schwarzschild black hole result is recovered in the limit $\xi\rightarrow0$, showing a consistently negative heat capacity and indicating a thermodynamically unstable solution. When corrections due to asymptotically safe gravity are taken into account ($\xi\neq 0$), the behavior of the heat capacity changes drastically. In particular, $C_V$ exhibits a divergence at a critical horizon radius, which coincides with the maximum of the Hawking temperature. This divergence signals a second-order phase transition. Moreover, for a finite range of $r_+$, the heat capacity becomes positive, indicating the emergence of a thermodynamically stable phase induced by quantum gravitational effects. We also observe that increasing the value of $\xi$ shifts the critical radius and smooths the thermodynamic behavior, reinforcing the interpretation of $\xi$ as a parameter controlling the strength of quantum corrections. These features are absent in the classical Schwarzschild solution and highlight the profound impact of asymptotically safe gravity on black hole thermodynamics.

\section{Topological analysis}\label{sV}

In this section we investigate the topological structure of the thermodynamic phase space associated with the improved Schwarzschild black hole. Previous thermodynamic studies of RG-improved Schwarzschild black holes have already revealed the existence of remnants, extrema of the Hawking temperature, and stable thermodynamic branches \cite{Mandal:2022quv}. However, to the best of our knowledge, these RG-improved solutions have not yet been examined within the framework of thermodynamic topology. This approach provides a powerful characterization of black hole phase space through topological charges associated with the generalized free energy and the inverse Hawking temperature. In the following, we first derive the topological defects directly from the exact off-shell free energy using the first law of black hole thermodynamics, and then discuss the global topological classification implied by the structure of the inverse Hawking temperature.

Here, the classification is performed using the generalized free energy, whose singularities and critical points encode the local topological structure of the phase space \cite{Wei:2022dzw,Wei:2024gfz}. Since the thermodynamic quantities of the RG-improved Schwarzschild solution receive quantum gravitational corrections encoded in the asymptotic-safety parameter $\xi$, it is natural to ask whether these modifications affect only the local thermodynamic behavior or also change the global topological structure of the thermodynamic phase space. Thermodynamic topology provides an appropriate framework to address this question by associating topological charges with the critical points of the generalized free energy and, consequently, allowing one to determine whether quantum gravitational effects preserve or modify the thermodynamic-topological class of the black hole.

The generalized free energy is defined as 
\begin{equation}\label{eq:free-energy-def}
\mathcal{F}(r_+)=M(r_+)-\frac{S(r_+)}{\tau},
\end{equation}
where $\tau$ is treated as an external control parameter. The on-shell thermodynamic free energy is recovered when $\tau^{-1}=T_H(r_+)$, while the off-shell extension allows for a topological classification of the thermodynamic phase space. Once the generalized free energy is determined, it can be used to define the auxiliary vector field $\vec{\phi}=(\phi^{r_+},\phi^{\Theta})$, whose components are given by
\begin{eqnarray}\label{eq:phi-components}
\phi^{r_+} &=& \frac{\partial \mathcal{F}}{\partial r_+}, \\
\phi^{\Theta} &=& -\cot{\Theta}\,\csc{\Theta}.
\end{eqnarray}
The zeros of the vector field $\vec{\phi}$ correspond to the critical points of the generalized free energy and define the thermodynamic states of the improved solution. These zeros can be interpreted as topological defects in the extended parameter space $(r_+,\Theta)$. Consequently, one can compute the associated winding numbers and analyze how quantum corrections arising from asymptotically safe gravity modify the global topological classification of the black hole solution.

The angular component vanishes identically at
\begin{equation}
\Theta^\star = \frac{\pi}{2},
\end{equation}
independently of the specific form of the free energy. Therefore, the nontrivial information about the critical behavior is entirely contained in the radial component $\phi^{r_+}$.

Although a closed analytic expression for the entropy is not available, the topological analysis does not require the explicit evaluation of the entropy itself. Instead, the radial component of the auxiliary vector field can be obtained directly from the first law of black hole thermodynamics. Indeed, from
\begin{equation}
dM=T_H\,dS,
\end{equation}
one immediately finds
\begin{equation}
S'(r_+)=\frac{M'(r_+)}{T_H(r_+)}.
\end{equation}
Since the generalized off-shell free energy is defined as
\begin{equation}
F(r_+)=M(r_+)-\frac{S(r_+)}{\tau},
\end{equation}
its radial derivative can be written exactly as
\begin{equation}
\phi^{r_+}=\frac{\partial F}{\partial r_+}=M'(r_+)-\frac{S'(r_+)}{\tau}=M'(r_+)\left(1-\frac{1}{\tau T_H(r_+)}
\right).
\label{eq:phi_exact}
\end{equation}
Therefore, the radial component of the vector field is determined entirely by the mass function and the Hawking temperature, without requiring an explicit integration of the entropy. Substituting the exact expressions for the mass, Eq.\eqref{eq:mass}, and the Hawking temperature, Eq.\eqref{eq:temperature}, into expression in Eq.~\eqref{eq:phi_exact} yields an exact analytic expression for the radial component of the vector field. However, the resulting formula is algebraically cumbersome and provides little additional physical insight. Since the topological classification depends only on the zeros of $\phi^{r_+}$, it is more convenient to analyze the factorized form of the Eq.\eqref{eq:phi_exact}, from which the critical points follow immediately. Indeed, along the physical branch of the solution the mass is a monotonically increasing function of the horizon radius, that is, $M'(r_+)>0$, so that the zeros of the radial component are entirely determined by the second factor in Eq.(\ref{eq:phi_exact}). Therefore, the critical points of the generalized free energy satisfy the exact condition 
\begin{equation}
T_H(r_+)=\frac{1}{\tau},
\label{eq:critical-condition}
\end{equation}
which provides precisely the on-shell free energy. This relation establishes a direct geometrical interpretation of the topological defects. They are precisely the intersection points between the Hawking-temperature curve and the horizontal line defined by the cavity temperature $T_H=\tau^{-1}$. 

The qualitative behavior of the Hawking temperature has already been discussed in Sec.IV and is displayed in Fig.\ref{fig:temperature}. Starting from the remnant configuration, the temperature vanishes at a finite horizon radius, increases monotonically until reaching a unique maximum, and then decreases asymptotically to zero in the infrared regime. Consequently, the number of critical points depends exclusively on the position of the horizontal line $T_H=\tau^{-1}$ with respect to the maximum temperature. Therefore, three different situations may occur. If we define $\beta(r_+)\equiv 1/T_H(r_+)$, then for
\begin{equation}
\tau<\beta_{\mathrm{min}},
\end{equation}
or equivalently $1/\tau>T_{\mathrm{max}}$, the horizontal line lies entirely above the Hawking-temperature curve and no critical point exists. When
\begin{equation}
\tau=\beta_{\mathrm{min}},
\end{equation}
the line becomes tangent to the curve at its maximum, giving rise to a degenerate critical point. Finally, for
\begin{equation}
\tau>\beta_{\mathrm{min}},
\end{equation}
the equation (\ref{eq:critical-condition}) admits two distinct solutions. These correspond respectively to the small-black-hole branch, located on the increasing part of the temperature curve, and the large-black-hole branch, located on the decreasing branch. Therefore, whenever $0<\tau^{-1}<T_{max}$ the generalized free energy possesses two distinct critical points, corresponding to two topological defects in the thermodynamic phase space. It is worth emphasizing that the emergence of this additional defect is a direct consequence of the quantum-gravity corrections encoded in the asymptotically safe running. In contrast to the classical Schwarzschild solution, whose Hawking temperature decreases monotonically and therefore admits a unique solution of $T_H=\tau^{-1}$, the RG-improved solution develops a finite-temperature maximum, allowing the equation $T_H=\tau^{-1}$ to admit two distinct solutions.

The stability of these critical points follows directly from the second derivative of the generalized free energy. Differentiating Eq.(\ref{eq:phi_exact}) once more, we obtain
\begin{equation}
\frac{\partial^2F}{\partial r_+^2}=M''(r_+)\left(1-\frac{1}{\tau T_H}\right)
+\frac{M'(r_+)T_H'(r_+)}{\tau T_H^2(r_+)}.
\end{equation}
At the critical points $r_+=r_+^*$, Eq.(\ref{eq:critical-condition}) implies
\begin{equation}
1-\frac{1}{\tau T_H}=0,
\end{equation}
so that the first term vanishes identically. Therefore,
\begin{equation}
\left.\frac{\partial^2F}{\partial r_+^2}\right|_{r_+^\star}=\frac{M'(r_+^\star)T_H'(r_+^\star)}{\tau T_H^2(r_+^\star)}.
\label{eq:F-second}
\end{equation}
Since $M'(r_+)>0$ along the physical branch and the denominator of Eq.~(\ref{eq:F-second}) is strictly positive, the sign of the second derivative is entirely determined by the slope of the Hawking temperature,
\begin{equation}
\mathrm{sign}\left(\frac{\partial^2F}{\partial r_+^2}\right)=\mathrm{sign} \left(T_H'(r_+)\right).
\end{equation}
Consequently, the critical point located on the increasing branch of the Hawking-temperature curve satisfies $T_H'(r_+)>0$, implying
\begin{equation}
\frac{\partial^2F}{\partial r_+^2}>0,
\end{equation}
and therefore corresponds to a local minimum of the generalized free energy. Conversely, the critical point located on the decreasing branch satisfies $T_H'(r_+)<0$, leading to
\begin{equation}
\frac{\partial^2F}{\partial r_+^2}<0,
\end{equation}
which characterizes a local maximum of the generalized free energy. Having established the stability of the critical points, we now determine the associated topological charges. 

Within the framework of thermodynamic topology, each isolated zero of the auxiliary vector field $\vec{\phi}$ is characterized by an integer-valued winding number $w$, which describes the local topology around the defect. The global topological charge of the black hole is then obtained by summing the winding numbers associated with all isolated defects,
\begin{equation}
W=\sum_i w_i,
\end{equation}
where the sum runs over all critical points of the generalized free energy.
Following the prescription introduced in Ref.~\cite{Wei:2022dzw}, the winding number is determined by the sign of the Jacobian of the vector field evaluated at each isolated defect. The Jacobian is given by
\begin{equation}
\det J=\left.\frac{\partial\phi^{r_+}}{\partial r_+}\right|_{r_+^\star}\,
\left.\frac{\partial\phi^\Theta}{\partial\Theta}\right|_{\Theta=\pi/2},
\end{equation}
where $r_+^\star$ denotes a critical point satisfying $\phi^{r_+}(r_+^\star)=0$. 

An equivalent computation can also be performed using the residue method developed in Ref.\cite{Fang:2022rsb}, where the winding number is extracted from the pole structure of the inverse radial component of the vector field. In both formulations, the winding number is ultimately determined by the sign of the second derivative of the generalized free energy evaluated at the critical point. However, as shown in Ref.\cite{Silva:2025iip}, this reduction is valid provided that the combination
\begin{equation}
M''(r_+)S'(r_+)-M'(r_+)S''(r_+)\neq0
\label{eq:regularity_condition}
\end{equation}
holds at the critical point, ensuring that the zero of the vector field is non-degenerate and that the thermodynamic mapping between the off-shell free energy and the $(M,S)$ representation remains regular. Using the first law of black hole thermodynamics,
\begin{equation}
S'(r_+)=\frac{M'(r_+)}{T_H(r_+)},
\end{equation}
we obtain
\begin{equation}
S''(r_+)=\frac{M''(r_+)}{T_H(r_+)}-\frac{M'(r_+)T_H'(r_+)}{T_H^2(r_+)}.
\end{equation}
Substituting these expressions into Eq.(\ref{eq:regularity_condition}) yields
\begin{equation}
M''S'-M'S''=\frac{\left(M'(r_+)\right)^2T_H'(r_+)}{T_H^2(r_+)}.
\end{equation}
Since $M'(r_+)>0$ along the physical branch and $T_H(r_+)>0$ for all physical black holes, this quantity vanishes only if $T_H'(r_+)=0$, corresponding to the degenerate case where the cavity temperature is tangent to the Hawking-temperature curve. Therefore, for the non-degenerate critical points satisfying
\begin{equation}
0<\tau^{-1}<T_{\rm max},
\end{equation}
one has
\begin{equation}
M''S'-M'S''\neq0,
\end{equation}
and the regularity condition of Ref.\cite{Silva:2025iip} is automatically satisfied. Consequently, the topological classification presented here is fully consistent with both the Jacobian and residue formulations.

For the angular component of the Jacobian,
\begin{equation}
\phi^\Theta=-\cot\Theta\,\csc\Theta,
\end{equation}
one finds
\begin{equation}
\left.\frac{\partial\phi^\Theta}{\partial\Theta}\right|_{\Theta=\pi/2}=1>0.
\end{equation}
Therefore, the angular contribution preserves the sign of the Jacobian, implying that
\begin{equation}
\mathrm{sign}(\det J)=\mathrm{sign}\left(\frac{\partial^2F}{\partial r_+^2}\right).
\end{equation} 

Following the thermodynamic topology framework of Ref.\cite{Wu:2024asq}, the local topological charge associated with each defect is identified with its winding number. Since the critical point on the increasing branch of the Hawking-temperature curve corresponds to a local minimum of the generalized free energy, its Jacobian is positive and the associated winding number is $w=+1$. Conversely, the critical point on the decreasing branch corresponds to a local maximum of the generalized free energy. Its Jacobian is therefore negative, yielding $w=-1$. The global topological charge is obtained by summing the winding numbers of all isolated defects, providing
\begin{equation}
W=\sum_i w_i
=(+1)+(-1)=0.
\end{equation}

This result should be contrasted with the classical Schwarzschild black hole. In the absence of asymptotically safe quantum-gravity corrections, the Hawking temperature is a monotonically decreasing function of the horizon radius. Consequently, the equation $T_H=\tau^{-1}$ admits a unique solution, giving rise to a single thermodynamic defect with winding number $w=-1$, and therefore to the well-known global topological charge $W=-1$.

The situation changes qualitatively once the renormalization-group improvement is taken into account. The quantum-gravity corrections regularize the Hawking-temperature profile, producing a finite-temperature maximum and allowing the equation $T_H=\tau^{-1}$ to possess two distinct solutions whenever $0<\tau^{-1}<T_{\rm max}$. As a consequence, a second topological defect emerges on the stable small-black-hole branch, carrying the opposite winding number $w=+1$. The cancellation between the two local topological charges changes the global topological charge from $W=-1$ to
\begin{equation}
W=0,
\end{equation}
showing that asymptotically safe gravity modifies not only the local thermodynamic properties of the black hole but also the global topology of its thermodynamic phase space. Accordingly, within the thermodynamic-topology classification proposed in Ref.\cite{Wu:2024asq}, the RG-improved Schwarzschild black hole belongs to the $W^{0+}$ topological class, whereas the classical Schwarzschild solution belongs to the $W^{1-}$ class. The transition from the $W^{1-}$ to the $W^{0+}$ class thus provides a genuine topological signature of the asymptotic-safety approach.

Finally, since the generalized free energy relies on the specific functional form of the quantum corrections, one might naturally wonder whether alternative renormalization group (RG) schemes could alter the critical-point structure and modify the topological class. As demonstrated by the exact condition in Eq.\eqref{eq:critical-condition}, the number of topological defects is dictated exclusively by the qualitative geometric profile of the Hawking temperature curve. Some RG-improved regular geometries may suppress the ultraviolet temperature divergence and generate a zero-temperature remnant. While alternative RG schemes or regular black-hole models could, in principle, generate different temperature profiles, multiple extrema, or different remnant behaviors, the $W=0$ classification is expected to be robust for RG-improved models characterized by a single maximum in $T_H(r_+)$ and a zero-temperature remnant. In such scenarios, smoothly vanishing in the deep UV and recovering the classical monotonic behavior in the IR, the isothermal condition $T_H(r_+)=\tau^{-1}$ will persistently admit exactly two roots for any cavity temperature in the range $0<\tau^{-1}<T_{\rm max}$. Therefore, the presence of two topological defects with opposite winding numbers, yielding the global topological charge $W=0$, remains a robust macroscopic feature within this broad class of regularized black holes, largely independent of the microscopic details of specific RG trajectories.

\section{Conclusions}\label{sVI}

We have constructed an explicit renormalization–group improved Schwarzschild geometry using the exact scheme-B running of the Newton coupling in the $\Lambda_0=0$ limit together with the standard interpolating proper-distance scale identification. This procedure yields a fully analytic coordinate-dependent coupling $G(r)$ and a closed expression for the improved lapse function.

The solution consistently interpolates between the infrared and ultraviolet regimes. At large distances, the classical Schwarzschild metric is recovered with subleading corrections of order $r^{-3}$. In the ultraviolet, within the physically admissible parameter range $\gamma>0$, the lapse function develops a de Sitter–like core and the Kretschmann scalar remains finite at the origin, demonstrating how asymptotically safe running effects regularize the curvature singularity within this effective geometry.

The parameter $\xi$ controls the strength of quantum corrections and determines the horizon structure. A critical curve separates black hole configurations from horizonless geometries, preventing the formation of arbitrarily small black holes. Thermodynamically, the Hawking temperature exhibits a maximum and vanishes at finite radius, while the heat capacity develops a divergence signaling a second-order phase transition and the emergence of a locally stable phase. 

Although a closed analytic expression for the entropy is not available, a controlled expansion in $\xi$ reveals a logarithmic correction to the area law, a feature commonly encountered in several other quantum gravity approaches. The topological analysis showed that asymptotically safe gravity modifies not only the local thermodynamic properties of the Schwarzschild black hole but also the global topology of its thermodynamic phase space. The regularization of the Hawking-temperature curve leads to the emergence of an additional stable topological defect with winding number ($w=+1$), which cancels the classical unstable contribution ($w=-1$). Consequently, the global topological charge changes from ($W=-1$) to ($W=0$), identifying the RG-improved Schwarzschild black hole as a member of the ($W^{0+}$) thermodynamic-topological class.

Finally, within the asymptotically safe gravity program, the cutoff scale parameter $\xi$ is typically expected to be of the order of the Planck scale, confining the strongest deviations from classical relativity to the deep ultraviolet regime. Consequently, the emergence of a zero-temperature state at a finite horizon radius provides a natural mechanism for the formation of Planck-size cold remnants, which are frequently discussed in the literature as potential dark matter candidates or as a resolution to the black hole information paradox. Furthermore, bridging this theoretical framework with current observational constraints is a highly active front. Indeed, following the introduction of the exact regularized geometry proposed in this work, its phenomenological implications have already begun to be explored in the recent literature \cite{Al-Badawi:2026zav,Yang:2026lbj,Borah:2026sja}. This includes the analysis of black hole shadows, ringdown profiles, and strong cosmic censorship, the impact of plasma on shadow radii and gravitational weak lensing constraints, as well as the computation of quasinormal modes under scalar perturbations. These recent developments suggest that our metric may serve as a useful phenomenological background for future tests of quantum gravitational effects against Event Horizon Telescope (EHT) imaging and future gravitational-wave data.

\acknowledgments{We acknowledge the financial support provided by the Conselho Nacional de Desenvolvimento Científicoe Tecnológico (CNPq), Fundação Cearense de Apoio ao Desenvolvimento Científico e
Tecnológico (FUNCAP) and Coordena\c c\~{a}o de Aperfei\c coamento de Pessoal de N\'{i}vel Superior - Brasil (CAPES) - Finance Code 001.
}
\bibliographystyle{apsrev4-1}
\bibliography{ref.bib}
\end{document}